\def\be{\begin{equation}}
\def\ee{\end{equation}}
\def\ba{\begin{eqnarray}}
\def\ea{\end{eqnarray}}
\def\l{\label}

\documentstyle[12pt]{article} \begin{document} 
\title{Measurement of entropy of a multiparticle system: a do-list }

\author{A.Bialas and W.Czyz\\ M.Smoluchowski Institute of Physics \\Jagellonian
University, Cracow\thanks{Address: Reymonta 4, 30-059 Krakow, Poland;
e-mail:bialas@trisc.if.uj.edu.pl; czyz@trisc.if.uj.edu.pl} \\and\\
H.Niewodniczanski Institute of Nuclear Physics, Cracow}
\maketitle

\begin{abstract}

An algorithm for measurement of entropy in multiparticle systems, based
on the recently published proposal of the present authors \cite{bc}, is
given.  Dependence on discretization of the system and effects of multiparticle 
correlations are discussed in some detail.   
\end{abstract}
 
It was suggested recently \cite{bcw} that studying event-by-event
fluctuations may be used for determination of entropy of multiparticle
systems created in high-energy collisions. A generalization of this idea 
and a specific proposal for  measurement of entropy  were formulated in
\cite{bc}. In the present note we spell out explicitly the steps to be
taken to implement effectively the method proposed in \cite{bc}. 
Importance of  the dependence of  measurements on discretization of
particle momenta and the role of (multi)particle correlations are
emphasized.

{\bf 1. Selection of the phase-space region.}

As the first step in the process of measurement one has to select a
phase-space region in which measurements are to be performed. This of
course depends on the detector acceptance as well as on the physics one
wants to investigate. The region cannot be too large because for large
systems the method is difficult to apply (the requirements on statistics
become too demanding). With the statistics of $10^6$ events, the region
containing (on the average) $\approx 100$ or less particles should be
possible to investigate. A reasonable procedure seems to be to start
from a small region and then increase it until the errors become
unbearable.

Comment 1: The proposed measurement is {\it not} restricted to systems
with very large number of particles. It can be applied to any
multiparticle system, e.g., in $e^+e^-$ annihilation, hadron-hadron
collisions or peripheral nucleus-nucleus collisions.

{\bf 2. Discretization of the spectrum. }

The selected phase-space region should now be divided into bins of equal
size in momentum space. The number of bins cannot be too large if one
wants to keep errors under control. On the other hand, as argued below,
it is important to study the dependence of results on the size (and thus the
number) of the bins. Therefore, large statistics is essential for a success
of the measurement.


Comment 2: If one chooses the bins which are not of equal size in momentum
space, the original expression for entropy requires a correction which
follows from an appropriate change of variables \cite{bc}. This
correction is, in general, not easy to calculate. Nevertheless it may be
interesting to study the dependence on the shape of the binning, as well.

{\bf 3. Description of an event.}

Using this procedure, an event is characterized by the number of
particles in each bin, i.e. by a set of integer numbers $s\equiv
m_i^{(j)}$, where $i=1,...M$ ($M$ is the total number of bins) and the
superscript $(j)$ runs over all kinds of particles present in the
final state. These sets represent different states of the
multiparticle system which were realized in a given experiment. The
number of possible different sets is, generally, very large (for 5 bins
and 100 particles one obtains $\approx 10^6 $ sets). This is, in fact,
the main difficulty in application of the proposed method. It simply
reflects the fact that the system we are dealing with has very many
states. 

Comment 3: It should be realized that, in practice, such a description is
never complete, i.e., it never describes fully the event. Most often
some of the variables are summed over. This is the case, e.g., when one
measures only charged particles. Then all the variables (i.e.
multiplicities and momenta) related to neutral particles are summed over.
It may be thus interesting to study {\it reduced events}, when even some
of the measured variables (e.g. particle identity) are summed over (i.e.
ignored).

{\bf 4.  Measurement of coincidence probabilities.}

As explained in \cite{bc}, this is the basis of the method and therefore
the most important step in the whole procedure.

The measurement consists of the simple counting how many times ($n_s$)
any given set $s$ appears in the whole sample of events\footnote{ Since
the number of different sets is very large, most of them shall appear
only once or not at all.}. Once the numbers $n_s$ are known for all
sets, one forms the sums:
\be
N_k= \sum_s n_s(n_s-1)...(n_s-k+1)   ,  \l{1}
\ee
with $k=1,2,3,...$.  The sum formally 
runs over all sets s recorded in a given experiment,
but nonvanishing contributions give
 only those which were recorded at least $k$ times.
 One  sees that $N_k$
is the total number of observed coincidences of $k$ configurations. 
The {\it coincidence probability} of  $k$ configurations is thus given by
\be
C_k = \frac{N_k}{N(N-1)...(N-k+1)}    ,           \l{2}
\ee
where $N$ is the total number of the events in the sample\footnote{As
explained in \cite{bc} this ratio is equal to the $(k-1)$-th moment of the
probability distribution $C_k = \sum_s (p_s)^k $. The proof follows
closely the argument of \cite {bp}.}.
  
One sees that $N_k$ given by (\ref{1}) are simply factorial moments of
the distribution of $n_s$ \cite{bp}. It is also clear that, since
$\sum_s n_s =N$, $C_1=1$. Finally, one sees that only states with $n_s
\geq k$ contribute to $N_k$ (and thus also to $C_k$).

{\bf 5. Errors.}

The error of $C_k$ is determined by the error of the numerator in
(\ref{2}). This error can be estimated by standard methods used in
evaluation of the moments of a distribution.

{\bf 6. Renyi entropies and Shannon entropy.} 

Once the coincidence probabilities $C_k$ $(k=1,2,...)$ are measured, it
is convenient \cite{bc} to calculate the
 Renyi entropies defined by \cite{r}
\be
H_k\equiv - \frac{\log C_k}{k-1}     .  \l{4}
\ee
The Shannon entropy $S$ (i.e. the standard statistical entropy) is formally
equal  to the limit of $H_k$ as $k\rightarrow 1$ and thus can 
only be obtained by extrapolation from a
series of measured values: $H_k=H_2,H_3,....$ to $k=1$ \footnote{
Obviously, one cannot just put $k=1$ in the formula (\ref{4}) 
 for that purpose: since
$C_1 = 1$, the R.H.S. of (\ref{4}) for $k=1$ represents the undefined
symbol $0/0$. }. Of course such an extrapolation procedure is not unique
and introduces uncertainty. The main point is, as usual, to choose the
"best" extrapolation formula, i.e. the functional dependence of $H_k$ on
$k$ which will be used to reach the point $k=1$ from the measured points
$k=2,3,...$. This form can only be guessed on the basis of physics
arguments (or prejudicies). 

In \cite{bc} it was suggested to use
\be
H_k= a\frac {\log k}{k-1} + a_0+ a_1(k-1) +a_2(k-1)^2 +....\;\;\;, \l{5}
\ee
where the number of terms is determined by the number of measured Renyi
entropies. This formula turned out to be very effective in reproducing the
correct value of entropy for some typical distributions encountered in
high-energy collisions. 

Another possibility is to use
\be
H_k = a_0 + a_1/k + a_2/k^2 +....  \l{6}
\ee
suggested by the formula for the free gas of massless
bosons\footnote{For the free gas of massless bosons the Renyi entropies
are given by $H_k=(1+1/k+1/k^2+1/k^3)S/4$ where $S$ is the Shannon
entropy.}. 

It will be interesting to compare the results from these two formulae.

Comment 4: The measured values of the Renyi entropies give valuable
information about the system and thus are of great interest,
independently of the accuracy of the extrapolation.

{\bf 7. Dependence on discretization; Scaling.}

As the result of the procedure explained in Sections 1 to 6, we obtain the 
Renyi entropies $H_k$, (k=2,3,...) and the Shannon entropy $S$ of a
given phase-space region. These entropies still depend on the method of
discretization of the momentum spectrum, in particular on the size of
the binning. If the bins are small enough and if the system is close to
thermal equilibrium (i.e. if fluctuations are small), one expects
 the following scaling law to hold
\be 
H_k(lM) = H_k(M) +\log l \;\;\; \rightarrow \;\;\;
S(lM) = S(M) + \log l \l{7} 
\ee 
 ($M$ and $lM$ are numbers of bins in two different
discretizations). If the scaling law is verified, one can determine the part
of entropy which is independent of binning. 

The rule (\ref{7}) is not expected to hold if the system is far from
thermal equilibrium and the fluctuations of the particle distribution
are large. In particular, the effects of intermittency \cite{bp} and
erraticity \cite{hwa} as implied, e.g., by a cascading mechanism of
particle production are expected to violate (\ref{7}). Thus testing the
dependence of entropies on the number of bins may reveal interesting
features of the system.

{\bf 8. Comparison of different regions; Additivity.}

Measurements of the  entropies $H_k$ and  $S$,
 as described above, can be performed independently
(and - in fact- simultaneously) in different phase-space regions. 
The results should give information on the entropy density and its possible
dependence on the position in phase-space (e.g., it seems likely that
the results in the central rapidity region may be rather different from
those in the projectile or target fragmentation). Furthermore, it is
important to verify to what  an extent the obtained entropies are additive,
i.e., whether the entropies measured in a region  $R$ which is the sum of
two regions $R_1$ and $R_2$ satisfy
\be
H_k(R) =H_k(R_1)+H_k(R_2) \;\;\; \rightarrow \;\;\; 
S(R)=S(R_1)+S(R_2).  \l{8}
\ee
 Eq. (\ref{8}) should be satisfied if there are no strong
correlations between the particles belonging to the regions  $R_1$ and
$R_2$. Thus verification of (\ref{8}) gives information about the
correlations between different phase-space regions.

Comment 5: It may be worth to point out that the scaling law (\ref{7})
and the additivity (\ref{8}) can be more precisely tested for Renyi
entropies ($H_k$) than for the Shannon entropy ($S$) where the
extrapolation procedure (described in Section 6) introduces always an
additional uncertainty.

{\bf 9. Conclusions}

 In conclusion, one sees that the measurement of entropy in
limited regions of phase-space is feasible. Moreover, even the  simplest
tests of the general scaling and additivity rules can provide essential
information on fluctuations and on correlations in the system.
It should be emphasized that for these tests the Renyi entropies turn
out to be  more useful than the standard Shannon entropy.

\vspace{0.3cm}

{\bf Acknowledgements} We thank Y.Foka for suggesting preparation of 
this note, to K.Fialkowski, A.Ostruszka and J.Wosiek for 
discussions and M.Gazdzicki for correspondence.

\vspace{0.3cm}

This investigation was supported in part by the KBN Grant No 2 P03B 086
14 and by Subsydium FNP 1/99.

\end{document}